\documentclass[aps,amsmath,preprintnumbers,amssymb,twocolumn,superscriptaddress,longbibliography,nofootinbib,showpacs]{revtex4-1}

\usepackage{graphicx}
\usepackage{dcolumn}
\usepackage{bm}
\usepackage{color}
\usepackage{comment}
\usepackage{float}
\usepackage[normalem]{ulem}
\usepackage[colorlinks=true,pdfstartview=FitV,linkcolor=blue,citecolor=blue,urlcolor=blue]{hyperref}
\usepackage[mathlines]{lineno}

\everymath{\displaystyle}
\begin{document}
\preprint{FERMILAB-PUB-23-256-PPD}

\newcommand{\up}[1]{$^{#1}$}
\newcommand{\down}[1]{$_{#1}$}
\newcommand{\powero}[1]{\mbox{10$^{#1}$}}
\newcommand{\powert}[2]{\mbox{#2$\times$10$^{#1}$}}

\newcommand{\mchi}{\mbox{$m_\chi$}}
\newcommand{\dedx}{$\rm{dE/dx}$}
\newcommand{\gev}{\mbox{GeV\,}$c^{-2}$}
\newcommand{\swn}{\mbox{$\sigma_{\chi-n}$}}
\newcommand{\evr}{\mbox{eV$_{\rm nr}$}}
\newcommand{\eve}{\mbox{eV$_{\rm ee}$}}
\newcommand{\dru}{\mbox{keV$_{\rm ee}^{-1}$\,kg$^{-1}$\,day$^{-1}$}}
\newcommand{\um}{\mbox{$\mu$m}}
\newcommand{\sxy}{\mbox{$\sigma_{xy}$}}
\newcommand{\sx}{\mbox{$\sigma_{x}$}}
\newcommand{\smax}{\mbox{$\sigma_{\rm max}$}}
\newcommand{\spix}{\mbox{$\sigma_{\rm pix}$}}
\newcommand{\obo}{\mbox{1$\times$1}}
\newcommand{\obh}{\mbox{1$\times$100}}
\newcommand{\dll}{\mbox{$\Delta LL$}}
\newcommand*\diff{\mathop{}\!\mathrm{d}}
\newcommand*\Diff[1]{\mathop{}\!\mathrm{d^#1}}

\newcommand{\kgd}{kg-day}
\newcommand{\tritium}{\mbox{$^{3}$H}}
\newcommand{\ironfive}{\mbox{$^{55}$Fe}}
\newcommand{\coseven}{\mbox{$^{57}$Co}}
\newcommand{\pbten}{$^{210}$Pb}
\newcommand{\biten}{$^{210}$Bi}
\newcommand{\sitwo}{$^{32}$Si}
\newcommand{\ptwo}{$^{32}$P}
\newcommand{\Ez}{\mbox{$E\text{-}z$} }
\newcommand{\Esigma}{\mbox{$E\text{-}\sigma_x$} }

\title{Confirmation of the spectral excess in DAMIC at SNOLAB with skipper CCDs}

\author{A.\,Aguilar-Arevalo}
\affiliation{Universidad Nacional Aut{\'o}noma de M{\'e}xico, Mexico City, Mexico}

\author{I.\,Arnquist}
\affiliation{Pacific Northwest National Laboratory (PNNL), Richland, WA, United States}

\author{N.\,Avalos}
\affiliation{Centro At\'{o}mico Bariloche and Instituto Balseiro, Comisi\'{o}n Nacional de Energ\'{i}a At\'{o}mica (CNEA), Consejo Nacional de Investigaciones Cient\'{i}ficas y T\'{e}cnicas (CONICET), Universidad Nacional de Cuyo (UNCUYO), San Carlos de Bariloche, Argentina}

\author{L.\,Barak}
\affiliation{School of Physics and Astronomy, 
 Tel-Aviv University, Tel-Aviv, Israel}

\author{D.\,Baxter}
\affiliation{Fermi National Accelerator Laboratory, Batavia, IL, United States}

\author{X.\,Bertou}
\affiliation{Centro At\'{o}mico Bariloche and Instituto Balseiro, Comisi\'{o}n Nacional de Energ\'{i}a At\'{o}mica (CNEA), Consejo Nacional de Investigaciones Cient\'{i}ficas y T\'{e}cnicas (CONICET), Universidad Nacional de Cuyo (UNCUYO), San Carlos de Bariloche, Argentina}

\author{I.M.\,Bloch}
\affiliation{Berkeley Center for Theoretical Physics, University of California, Berkeley, CA, United States}
\affiliation{Lawrence Berkeley National Laboratory, Berkeley, CA, United States}

\author{A.M.\,Botti}
\affiliation{Fermi National Accelerator Laboratory, Batavia, IL, United States}

\author{M.\,Cababie}
\affiliation{Universidad de Buenos Aires, Facultad de Ciencias Exactas y Naturales, Departamento de F\'isica, Buenos Aires, Argentina}
\affiliation{CONICET - Universidad de Buenos Aires, Instituto de F\'isica de Buenos Aires (IFIBA), Buenos Aires, Argentina}
\affiliation{Fermi National Accelerator Laboratory, Batavia, IL, United States}

\author{G.\,Cancelo}
\affiliation{Fermi National Accelerator Laboratory, Batavia, IL, United States}

\author{N.\,Castell\'{o}-Mor}
\affiliation{Instituto de F\'{i}sica de Cantabria (IFCA), CSIC - Universidad de Cantabria, Santander, Spain}

\author{B.A.\,Cervantes-Vergara}
\affiliation{Universidad Nacional Aut{\'o}noma de M{\'e}xico, Mexico City, Mexico} 

\author{A.E.\,Chavarria}
\affiliation{Center for Experimental Nuclear Physics and Astrophysics, University of Washington, Seattle, WA, United States}

\author{J.\,Cortabitarte-Guti\'{e}rrez}
\affiliation{Instituto de F\'{i}sica de Cantabria (IFCA), CSIC - Universidad de Cantabria, Santander, Spain}

\author{M.\,Crisler}
\affiliation{Fermi National Accelerator Laboratory, Batavia, IL, United States}

\author{J.\,Cuevas-Zepeda}
\affiliation{Kavli Institute for Cosmological Physics and The Enrico Fermi Institute, The University of Chicago, Chicago, IL, United States}

\author{A.\,Dastgheibi-Fard}
\affiliation{LPSC LSM, CNRS/IN2P3, Universit\'{e} Grenoble-Alpes, Grenoble, France}

\author{C.\,De Dominicis}
\affiliation{Laboratoire de physique nucl\'{e}aire et des hautes \'{e}nergies (LPNHE), Sorbonne Universit\'{e}, Universit\'{e} Paris Cit\'{e}, CNRS/IN2P3, Paris, France}

\author{O.\,Deligny}
\affiliation{CNRS/IN2P3, IJCLab, Universit\'{e} Paris-Saclay, Orsay, France}

\author{A.\,Drlica-Wagner}
\affiliation{Fermi National Accelerator Laboratory, Batavia, IL, United States}
\affiliation{Kavli Institute for Cosmological Physics and The Enrico Fermi Institute, The University of Chicago, Chicago, IL, United States}
\affiliation{Department of Astronomy and Astrophysics, University of Chicago, Chicago, IL, United States}

\author{J.\,Duarte-Campderros}
\affiliation{Instituto de F\'{i}sica de Cantabria (IFCA), CSIC - Universidad de Cantabria, Santander, Spain}

\author{J.C.\,D'Olivo}
\affiliation{Universidad Nacional Aut{\'o}noma de M{\'e}xico, Mexico City, Mexico} 

\author{R.\,Essig}
\affiliation{C.N.~Yang Institute for Theoretical Physics, Stony Brook University, Stony Brook, NY, United States}

\author{E.\,Estrada}
\affiliation{Centro At\'{o}mico Bariloche and Instituto Balseiro, Comisi\'{o}n Nacional de Energ\'{i}a At\'{o}mica (CNEA), Consejo Nacional de Investigaciones Cient\'{i}ficas y T\'{e}cnicas (CONICET), Universidad Nacional de Cuyo (UNCUYO), San Carlos de Bariloche, Argentina}

\author{J.\,Estrada}
\affiliation{Fermi National Accelerator Laboratory, Batavia, IL, United States}

\author{E.\,Etzion}
\affiliation{School of Physics and Astronomy, 
 Tel-Aviv University, Tel-Aviv, Israel}

\author{F.\,Favela-Perez}
\affiliation{Universidad Nacional Aut{\'o}noma de M{\'e}xico, Mexico City, Mexico} 

\author{N.\,Gadola}
\affiliation{Universit\"{a}t Z\"{u}rich Physik Institut, Z\"{u}rich, Switzerland}

\author{R.\,Ga\"{i}or}
\affiliation{Laboratoire de physique nucl\'{e}aire et des hautes \'{e}nergies (LPNHE), Sorbonne Universit\'{e}, Universit\'{e} Paris Cit\'{e}, CNRS/IN2P3, Paris, France}

\author{S.E.\,Holland}
\affiliation{Lawrence Berkeley National Laboratory, Berkeley, CA, United States}

\author{T.\,Hossbach}
\affiliation{Pacific Northwest National Laboratory (PNNL), Richland, WA, United States} 

\author{L.\,Iddir}
\affiliation{Laboratoire de physique nucl\'{e}aire et des hautes \'{e}nergies (LPNHE), Sorbonne Universit\'{e}, Universit\'{e} Paris Cit\'{e}, CNRS/IN2P3, Paris, France}

\author{B.\,Kilminster}
\affiliation{Universit\"{a}t Z\"{u}rich Physik Institut, Z\"{u}rich, Switzerland}

\author{Y.\,Korn}
\affiliation{School of Physics and Astronomy, 
 Tel-Aviv University, Tel-Aviv, Israel}

\author{A.\,Lantero-Barreda}
\affiliation{Instituto de F\'{i}sica de Cantabria (IFCA), CSIC - Universidad de Cantabria, Santander, Spain}

\author{I.\,Lawson}
\affiliation{SNOLAB, Lively, ON, Canada}

\author{S.\,Lee}
\affiliation{Universit\"{a}t Z\"{u}rich Physik Institut, Z\"{u}rich, Switzerland}

\author{A.\,Letessier-Selvon}
\affiliation{Laboratoire de physique nucl\'{e}aire et des hautes \'{e}nergies (LPNHE), Sorbonne Universit\'{e}, Universit\'{e} Paris Cit\'{e}, CNRS/IN2P3, Paris, France}

\author{P.\,Loaiza}
\affiliation{CNRS/IN2P3, IJCLab, Universit\'{e} Paris-Saclay, Orsay, France}

\author{A.\,Lopez-Virto}
\affiliation{Instituto de F\'{i}sica de Cantabria (IFCA), CSIC - Universidad de Cantabria, Santander, Spain}

\author{S.\,Luoma}
\affiliation{SNOLAB, Lively, ON, Canada}

\author{E.\,Marrufo-Villalpando}
\affiliation{Kavli Institute for Cosmological Physics and The Enrico Fermi Institute, The University of Chicago, Chicago, IL, United States}

\author{K.J.\,McGuire}
\affiliation{Center for Experimental Nuclear Physics and Astrophysics, University of Washington, Seattle, WA, United States}

\author{G.F.\,Moroni}
\affiliation{Fermi National Accelerator Laboratory, Batavia, IL, United States}

\author{S.\,Munagavalasa}
\affiliation{Kavli Institute for Cosmological Physics and The Enrico Fermi Institute, The University of Chicago, Chicago, IL, United States}

\author{D.\,Norcini}
\affiliation{Kavli Institute for Cosmological Physics and The Enrico Fermi Institute, The University of Chicago, Chicago, IL, United States}

\author{A.\,Orly}
\affiliation{School of Physics and Astronomy, 
 Tel-Aviv University, Tel-Aviv, Israel}

\author{G.\,Papadopoulos}
\affiliation{Laboratoire de physique nucl\'{e}aire et des hautes \'{e}nergies (LPNHE), Sorbonne Universit\'{e}, Universit\'{e} Paris Cit\'{e}, CNRS/IN2P3, Paris, France}

\author{S.\,Paul}
\affiliation{Kavli Institute for Cosmological Physics and The Enrico Fermi Institute, The University of Chicago, Chicago, IL, United States}

\author{S.E.\,Perez}
\affiliation{Universidad de Buenos Aires, Facultad de Ciencias Exactas y Naturales, Departamento de F\'isica, Buenos Aires, Argentina}
\affiliation{CONICET - Universidad de Buenos Aires, Instituto de F\'isica de Buenos Aires (IFIBA), Buenos Aires, Argentina}
\affiliation{Fermi National Accelerator Laboratory, Batavia, IL, United States}

\author{A.\,Piers}
\affiliation{Center for Experimental Nuclear Physics and Astrophysics, University of Washington, Seattle, WA, United States}

\author{P.\,Privitera}
\affiliation{Kavli Institute for Cosmological Physics and The Enrico Fermi Institute, The University of Chicago, Chicago, IL, United States}
\affiliation{Laboratoire de physique nucl\'{e}aire et des hautes \'{e}nergies (LPNHE), Sorbonne Universit\'{e}, Universit\'{e} Paris Cit\'{e}, CNRS/IN2P3, Paris, France}

\author{P.\,Robmann}
\affiliation{Universit\"{a}t Z\"{u}rich Physik Institut, Z\"{u}rich, Switzerland}

\author{D.\,Rodrigues}
\affiliation{Universidad de Buenos Aires, Facultad de Ciencias Exactas y Naturales, Departamento de F\'isica, Buenos Aires, Argentina}
\affiliation{CONICET - Universidad de Buenos Aires, Instituto de F\'isica de Buenos Aires (IFIBA), Buenos Aires, Argentina}
\affiliation{Fermi National Accelerator Laboratory, Batavia, IL, United States}

\author{N.A.\,Saffold}
\affiliation{Fermi National Accelerator Laboratory, Batavia, IL, United States}

\author{S.\,Scorza}
\affiliation{SNOLAB, Lively, ON, Canada}

\author{M.\,Settimo}
\affiliation{SUBATECH, Nantes Universit\'{e}, IMT Atlantique, CNRS-IN2P3, Nantes, France}

\author{A.\,Singal}
\affiliation{C.N.~Yang Institute for Theoretical Physics, Stony Brook University, Stony Brook, NY, United States}
\affiliation{Department of Physics and Astronomy, Stony Brook University, Stony Brook, NY, United States}

\author{R.\,Smida}
\affiliation{Kavli Institute for Cosmological Physics and The Enrico Fermi Institute, The University of Chicago, Chicago, IL, United States}

\author{M.\,Sofo-Haro}
\affiliation{Fermi National Accelerator Laboratory, Batavia, IL, United States}
\affiliation{Universidad Nacional de C\'ordoba, IFEG (CONICET) \& RA0 (CNEA), C\'ordoba, Argentina}

\author{L.\,Stefanazzi}
\affiliation{Fermi National Accelerator Laboratory, Batavia, IL, United States}

\author{K.\,Stifter}
\affiliation{Fermi National Accelerator Laboratory, Batavia, IL, United States}

\author{J.\,Tiffenberg}
\affiliation{Fermi National Accelerator Laboratory, Batavia, IL, United States}

\author{M.\,Traina}
\affiliation{Center for Experimental Nuclear Physics and Astrophysics, University of Washington, Seattle, WA, United States}

\author{S.\,Uemura}
\affiliation{Fermi National Accelerator Laboratory, Batavia, IL, United States}

\author{I.\,Vila}
\affiliation{Instituto de F\'{i}sica de Cantabria (IFCA), CSIC - Universidad de Cantabria, Santander, Spain}

\author{R.\,Vilar}
\affiliation{Instituto de F\'{i}sica de Cantabria (IFCA), CSIC - Universidad de Cantabria, Santander, Spain}

\author{T.\,Volansky}
\affiliation{School of Physics and Astronomy,   
 Tel-Aviv University, Tel-Aviv, Israel}

\author{G.\,Warot}
\affiliation{LPSC LSM, CNRS/IN2P3, Universit\'{e} Grenoble-Alpes, Grenoble, France}

\author{R.\,Yajur}
\affiliation{Kavli Institute for Cosmological Physics and The Enrico Fermi Institute, The University of Chicago, Chicago, IL, United States}

\author{T-T.\,Yu}
\affiliation{Department of Physics and Institute for Fundamental Science, University of Oregon, Eugene, OR, United States}

\author{J-P.\,Zopounidis}
\affiliation{Laboratoire de physique nucl\'{e}aire et des hautes \'{e}nergies (LPNHE), Sorbonne Universit\'{e}, Universit\'{e} Paris Cit\'{e}, CNRS/IN2P3, Paris, France}

\collaboration{DAMIC, DAMIC-M and SENSEI Collaborations}
\noaffiliation

\date{\today}

\begin{abstract}
We present results from a 3.25~\kgd~target exposure of two silicon charge-coupled devices (CCDs), each with 24 megapixels and skipper readout, deployed in the DAMIC setup at SNOLAB.
With a reduction in pixel readout noise of a factor of 10 relative to the previous detector, we investigate the excess population of low-energy events in the CCD bulk previously observed above expected backgrounds.
We address the dominant systematic uncertainty of the previous analysis through a depth fiducialization designed to reject surface backgrounds on the CCDs. 
The measured bulk ionization spectrum confirms the presence of an excess population of low-energy events in the CCD target with characteristic rate of ${\sim}7$ events per \kgd\, and electron-equivalent energies of ${\sim}80~$eV, whose origin remains unknown.
\end{abstract}


\maketitle

The DAMIC (DArk Matter In CCDs) experiment searches for the interaction of dark matter particles in the galactic halo~\cite{Kolb:1990vq} with silicon atoms in the fully depleted active region of scientific charge-coupled devices (CCDs).
Between 2017 and 2019, DAMIC acquired 11 kg-days of data with an array of seven CCDs with conventional readout (pixel noise \spix$\sim$1.6$~e^-$) installed in a low-background setup deep underground at SNOLAB~\cite{PhysRevLett.125.241803,*PhysRevD.105.062003}.
In December 2020, DAMIC reported a statistically significant ($3.7\sigma$) excess of events above its background model between the threshold energy of $50~\eve $ and $200~\eve$~\cite{PhysRevLett.125.241803,*PhysRevD.105.062003}.
The population of the excess events was best described by ionization events uniformly distributed in the bulk of the CCDs, with an exponentially decaying energy spectrum with decay energy $\varepsilon=67\pm37~\text{eV}_\text{ee}$ and a rate of $5.1\pm2.3$ events per kg-day.
The spectral fit was performed to all ionization events above the energy threshold in depth vs. energy space, with a full background model including different templates for bulk and surface events.
This approach was necessary because the bulk and surface populations could only be distinguished statistically, since the determination of the depth of the interaction was limited by readout noise.
Consequently, the dominant systematic uncertainty was identified as the modeling of surface backgrounds, and the statistically significant rise was reported only as an unmodeled excess of events over background.

In November 2021, the DAMIC setup was upgraded with two skipper CCDs, which can achieve single-electron resolution.
This article presents results from a 3.25 \kgd\ target exposure acquired throughout 2022 with the upgraded detector~\cite{alexthesis}.
The order-of-magnitude improvement in readout noise, from $\spix\sim1.6\,e^-$ to 0.16$\,e^-$, results in a significant improvement in the depth localization of ionization events. This allows for a new analysis based on a fiducial selection to obtain a clean sample of bulk events down to a threshold of 23\,\eve .
In addition, the lower threshold is expected to increase the measured rate of reconstructed excess events from 1.7 to 3.0~events per kg-day. 

The DAMIC skipper upgrade features a few notable differences from its predecessor~\cite{PhysRevLett.125.241803,*PhysRevD.105.062003}.
The previous seven DAMIC CCDs were decommissioned to install two 24-megapixel DAMIC-M skipper devices~\cite{PhysRevLett.130.171003} in a new oxygen-free high conductivity (OFHC) copper box.
The stock copper was stored in SNOLAB for five years to suppress cosmogenic activation~\cite{activationdoi:10.1142/S0217751X17430060, *baudisactive}, and brought to the surface at the time of machining.
The total sea-level exposure of the machined components is less than nine days.
The box with the two CCDs is connected to the existing cold finger to bring the sensors to a temperature ${\lesssim}140~$K.
The two ancient lead bricks employed in the previous installation are positioned above and below the CCD box to provide additional shielding from environmental backgrounds. 
CCD 1 is located at the bottom of the box, and CCD 2 at the top, with their front surfaces (on which the pixel array is patterned) oriented upward. There is a 2.5\,mm vertical gap between CCDs, with no material in between. 
The previous vacuum interface board (VIB) was replaced with a new one designed for skipper devices, which remains shielded by 18 cm of lead located above the CCDs~\cite{PhysRevLett.125.241803,*PhysRevD.105.062003}. 
CCD control and readout is performed with the Low Threshold Acquisition (LTA) electronics designed for SENSEI~\cite{8709274}.
In the setup, two LTA boards are synchronized on the same clock signal allowing low-noise readout of both CCDs.
Three-meter long coaxial cables carry the signals from the LTAs through the 42-cm polyethylene neutron shield to the air side of the VIB.

The 24-megapixel skipper CCDs were designed by Lawrence Berkeley National Laboratory (LBNL)~\cite{1185186} and fabricated by Teledyne/DALSA. The CCD substrate consists of $670~\mu\text{m}$ of fully depleted, high-resistivity (${>}10~\text{k}\Omega\cdot$cm) n-type silicon. Ionizing particles produce electron-hole pairs in the silicon as they lose energy. The charge carriers are drifted across the substrate, in the vertical direction ($\hat z$), by means of a bias voltage applied to a backside contact, $V_{\rm sub}=60$\,V.
Carriers experience thermal motion as they drift in the substrate electric field, which leads to a spatial variance in the transverse plane ($\hat x \hat y$) that is proportional to the transit time.
Holes are collected at a potential minimum created by the buried p-channel below the pixel array. 
The pixel array of each CCD consists of $6144\times4128$ (columns$\times$rows) pixels, each of size \mbox{$15~\mu\text{m}\times15~\mu\text{m}$}, for a total sensitive mass of $8.9~$g per CCD.
Clock signals are sent to the three-phase polysilicon gates on each pixel to transfer the collected charge across CCD rows ($\hat y$ direction) and into the horizontal register (bottom row), where it is transferred along $\hat x$ to two skipper amplifiers located at opposite ends (U and L) of the horizontal register. 
Skipper amplifiers can perform multiple nondestructive charge measurements (NDCMs) of the charge in a single pixel. Taking the average of $N_\text{sample}$ pixel samples improves the readout noise by a factor $1/\sqrt{N_\text{sample}}$, which enables single charge resolution with enough samples. 
The energy depositions from low-energy ionizing particles\textemdash whose tracks are much shorter than the pixel size\textemdash result in pixelated 2D Gaussian distributions of charge in the images. The lateral spread of the charge distribution, \sxy,  is positively correlated with the depth of interaction, with a maximum value $\smax \sim1~$pix.

Sources of single-electron events relevant in skipper CCDs have been characterized in Ref.~\cite{PhysRevApplied.17.014022}. DAMIC-M data at low charge multiplicities are consistent with a Poisson background from leakage current in the CCDs~\cite{PhysRevLett.130.171003}.
Uncorrelated single-electron events arise from a combination of instrumental and radiation backgrounds, and are hereafter referred to as dark counts. To minimize leakage current, CCD potentials were inverted to fill surface traps in several steps during cooldown, starting at 160~K until the final cold finger temperature of 110~K.
CCD biases were optimized to suppress light emission by the amplifiers, and the clock values were selected to minimize spurious charge~\cite{PhysRevApplied.17.014022} while maintaining high charge transfer efficiency.
Following commissioning and cooldown, the pixel readout noise for a single measurement ($N_{\rm sample}$=1) was ${\sim}4~e^-$ in all four skipper amplifiers. 
The background dark count rate in the images was between $2.4\times10^{-3}$ and $3.0\times10^{-3}~\text{e}^-/\text{pix}/\text{day}$.
The rate of accidental noise clusters, which sets the analysis threshold, depends on a combination of pixel readout noise, shot noise from dark counts, and the exposure per pixel. While increasing the signal integration time and the number of NDCMs decreases the readout noise, it also increases the overall readout time and, thus, the exposure per pixel.
An increase in the exposure per pixel increases the dark counts per pixel, and hence the shot noise. We performed studies on simulated images with the measured noise profile of the CCDs to establish the readout parameters.
For science data, CCDs were read out continuously with a signal integration time $t_\text{int}=8~\mu$s and $N_\text{sample}=460$.\footnote{Since the readout is continuous, the $y$ coordinate of the image pixel does not correspond to the $y$ coordinate of the event in the CCD pixel array.} This resulted in a pixel exposure time of $\sim$50 hours and a pixel noise $\spix \sim 0.16~e^-$.
The output data were stored as arrays of size \mbox{$3300 \times 210 \times460$} ($N_\text{columns}\times N_\text{rows}\times N_\text{sample}$), with one file per amplifier.
The CCD physical array is divided by two in the $\hat x$ direction since two amplifiers are used for readout. Each amplifier reads $3072$ physical pixels per row and an additional $228$ pixels past the physical extent of the array, which constitute the overscan.
Overscan pixels have significantly shorter exposure than physical ones, and allow a check on noise and charge transfer efficiency. 
Unlike previous DAMIC data runs, the charge was read out for every physical pixel, rather than after summing the charge from a group of pixels, since this provides better spatial resolution. 

A total exposure of 4.81~\kgd\ was acquired between February 2022 and January 2023 in seven data runs.
For the first run, in which a 1~\kgd\ exposure was collected, images with a smaller overscan ($N_\text{columns}=3100$) were acquired.
Images were processed as follows.
Images with pixel values in analog-to-digital units (ADUs) were constructed by averaging the value for each pixel across NDCMs. 
The first NDCM was excluded to avoid the noise transient at the start of pixel readout.
This procedure produces a \mbox{$3300\times 210$} (\mbox{$3100\times 210$}) average image.
The median pixel value is a good estimator of the baseline since $>96$\% of pixels have an occupancy of $0e^-$.
Thus, for every row in the average images, the median value was subtracted from every pixel value to correct for any baseline shift between rows. 
The distribution of averaged pixel values (PVD) exhibits discrete peaks corresponding to the pixel charge content ($0e^-$, $1e^-$, $2e^-$, etc.) that is dominated by dark counts.
The PVD of every image was fit to a Poisson distribution convolved with a Gaussian function to evaluate the gain (from the mean of the Gaussian, in ADU/$e^-$), noise (Gaussian standard deviation) and background dark counts (the Poisson mean).
A total of 10124 images were inspected during daily shifts and 142 were excluded based on several quality criteria, \emph{i.e.}, if the gain was outside the $3\sigma$ range defined for each amplifier and run, the pixel noise did not decrease with the expected $1/\sqrt{N_\text{sample}}$, or there was a prominent gradient in the baseline across the image.
Most of the discarded images were found to be correlated with electrical power instabilities at SNOLAB.
Finally, 40 average images were concatenated into a joint $3300\times 8400$ ($3100\times 8400$) image to reduce the number of ionization events split between images.

A mask was defined per amplifier to exclude defects and regions with increased dark counts.
Defects in the silicon lattice alter the local band gap, which leads to increased charge leakage at specific locations in the CCD~\cite{janesick2001scientific, *osti_1482433}. Because of continuous readout, all defects appear as ``hot'' columns in the science run images.
Any columns with $\lambda_i > 35\times10^{-3}~\text{e}^-/\text{pix}$ were excluded from the analysis since their PVD is typically not well described by a Poisson distribution, as estimated from a chi-square goodness-of-fit test.
Few columns with smaller $\lambda_i$ but exhibiting a non-Poisson charge distribution (p-value $<3.2 \times 10^{-5}$) were also excluded.
Columns with $x> 1565$ in the 1L amplifier (L amplifier of CCD 1) were discarded due to inefficient charge transfer past this point, possibly due to a charge trap in the horizontal register.
To maximize the sensitivity to defects, a set of images with $N_\text{sample}=1$ was acquired at a cold finger temperature of $160~$K, where charge leakage from defects is orders-of-magnitude larger. These ``warm'' images were read out after a 30-minute exposure of the pixel array, with defects appearing as localized hot pixels due to the relatively short readout time. We then generated a ``median image,'' where each pixel contains the median value of the pixel over all warm images, and identified defects as contiguous pixels $>5\sigma$ above the background noise.
In addition, we evaluated the median pixel value for every column over all images in the science data, and estimated a moving average of the column medians in a 200-column window.
Any column in the science data with median value 1.3 times ($2\sigma$) higher than the moving average and coincident with a defect in the warm images was excluded.
Finally, to reject events that are split between adjacent images, we masked a border of 10 pixels in the $\hat x$ direction and five pixels in the $\hat y$ direction around the joint image.
The amplifier masks account for a 52\,\%, 28\,\%, 12\,\% and 36\,\% loss in exposure in the 1L, 1U, 2L and 2U amplifiers, respectively.

To remove backgrounds associated with high-energy ionization events, pixels were also masked on a per-image basis.
Pixel clusters were identified as groups of contiguous pixels with charge $\geq 3~e^-$. The cluster energy was estimated as the sum of pixel values in the cluster assuming $3.8$~\eve$/e^-$~\cite{RODRIGUES2021165511}.
For every cluster with energy $>10$\,k\eve, we masked the smallest rectangular region of pixels containing the cluster plus a 2-pixel wide border around it.
Trails of charge from inefficient transfer of large charge packets from high-energy depositions were removed by excluding the 800 (100) trailing pixels in the horizontal direction in the L (U) images, and 20 trailing pixels in the vertical direction. These selections were obtained by studying the PVD in the trailing regions until there was no evidence of excess charge above background dark counts.
We observed $\mathcal O(10^{-4})$ cross talk between amplifiers, with high-energy events read out by one amplifier resulting in few-electron events in the other amplifiers.
While this is most relevant for $\alpha$ decays, we conservatively masked all pixels that were read out at the same time as pixels within the rectangular box surrounding high-energy clusters.
In addition, pixels for which the standard deviation of the NDCMs was larger than 10\,$e^-$ were excluded.
This selection accounts for occasional inaccurate estimates of the pixel charge due to noise fluctuations.
The per-image masks result in only a $0.6$\,\% loss in exposure with small variations per amplifier.
Overall, image selection and masking results in a final exposure of 3.25\,kg-day.

The cluster search was performed using a likelihood clustering algorithm on the images~\cite{PhysRevD.94.082006}. 
The likelihood $\mathcal{L}(q_{ij} | N, \vec{\mu}, \sigma_{xy}, \lambda_i, \sigma_{\rm pix})$ defines the probability that the pixel values $q_{ij}$ in a specified window are described by a 2D Gaussian distribution of charge with mean position $\vec{\mu}$, lateral spread \sxy, and amplitude $N$ on top of shot noise with mean occupancy $\lambda_i$, where every pixel has white noise \spix .
The images were scanned with a moving window of $5\times 5$ pixels, where, at every window position, we computed the likelihood that pixels come only from noise ($\mathcal L_n: N=0$) or noise plus a Gaussian charge distribution ($\mathcal L_g: N_0, \vec{\mu}_0, {\sigma_{xy}}_0$).
The initial guesses for $N_0$ and $\vec{\mu}_0$ are the total charge in the window and the center of the window, respectively. The initial guess for the lateral spread of the cluster ${\sigma_{xy}}_0$ was set to 1.0 pix. The values for $\lambda_i$ and \spix\ were fixed to the measured values for the specific image.
At window positions where there is a clear preference for a Gaussian cluster \mbox{($\ln\mathcal L_g - \ln\mathcal L_n > 7$)}, we vary the window position until \mbox{$\ln\mathcal L_g - \ln\mathcal L_n$} is maximized.
We then fix the window and perform a log-likelihood optimization with $N$, $\vec{\mu}$, and \sxy\ as free parameters.
The test statistic \mbox{$\Delta LL = - (\ln \mathcal{\tilde{L}}_g - \ln\mathcal L_n)$}, with $\mathcal{\tilde{L}}_g$ the best-fit likelihood, is a metric for the preference of the Gaussian hypothesis, with more negative values   
corresponding to a higher probability of ionization event within the selected window. 
The best-fit values for $N$, $\vec{\mu}$ and \sxy\ are taken as the best estimates for the energy, mean $(x,y)$ position, and lateral spread of the cluster. 
A second fit was performed with $\sigma_x$ and $\sigma_y$ as independent free parameters.

A selection on \dll\ was used to reject accidental clusters from noise.
We started by simulating ``blank'' images, which contain only the measured pixel readout noise and shot noise from dark counts as a function of column number in the data.
We ran the likelihood clustering on \mbox{30~kg-day} of simulated blanks to get a distribution of accidental noise clusters, which had a maximum charge of $10~e^-$.
We compared the distributions between data and blanks for clusters with $q\leq10~e^-$ and $-25 < \Delta LL < -10$, where the spectrum is dominated by noise clusters.
Excellent agreement in both cluster charge and the $x$ coordinate of the cluster center was found.
To obtain an accidental rate of \mbox{$R_\text{noise} < 0.01~$(\kgd)$^{-1}$} uniformly across the charge range $[5, 10]~e^-$, we select events with $\Delta LL \lesssim -29$, with small (at most $\pm 2$) variations between charge bins because of changes in the shape and amplitude of the $\Delta LL$ distributions.
Above $10~e^-$, a fixed selection $\Delta LL < -28$ was used.  

A series of criteria was applied to select ``valid'' clusters.
We select clusters whose fit window does not contain any masked pixels and which are properly centered ($\vec{\mu}<1.2$ from window center).
To select events with the expected 2D Gaussian topology, we first excluded extended clusters with $\sigma_x$ or $\sigma_y$ greater than 1.5~\smax.
Ionization events interacting in the field-free regions past the horizontal register can produce horizontal clusters with degraded energy~\cite{Moroni:2021lyt}.
Such asymmetric clusters were rejected by excluding events with $\sigma_y < 0.2$ and $\sigma_x > 0.5$ \mbox{pix}.
We applied an analogous rejection for vertical events ($\sigma_x < 0.2$ and $\sigma_y > 0.5$), which can originate from charge released by traps in the CCD pixel array.
Finally, to exclude clusters from trails of charge that were missed by the mask, we require that no more than 8\% of pixels within $\pm50$ pixels about the center of the cluster along the row had a value $\geq 1~e^-$.

The variance of the lateral diffusion of charges (the lateral spread of a low-energy event) can be modeled as~\cite{1185186,PhysRevLett.125.241803,*PhysRevD.105.062003}:
\begin{equation}
\label{eq:diffmodel}
\sigma^2_{xy}(z,E)=-A\ln|1-bz| \cdot (\alpha+\beta E)^2,
\end{equation}
where $z$ and $E$ denote event depth and energy, respectively, and parameters $A$ and $b$ depend on the microscopic properties of the substrate at operating temperature and $V_{\rm sub}$.
Best-fit values for the parameters \mbox{$A~=~3.07~$pix$^2$ and $b~=~5.35\times10^{-4}$~$\mu\text{m}^{-1}$} were obtained in a surface setup using straight cosmic muon tracks~\cite{PhysRevD.94.082006}.
The values of $\alpha = 0.889$ and $\beta = 7.4\times10^{-3}$~keV$_{\rm ee}^{-1}$ were obtained from a fit to backside events (predominantly \pbten\ and \biten\ decays) as a function of energy in the SNOLAB data.
We performed a selection on \sxy\ as a function of energy to remove 95.4\% of ionization events originating from the surfaces of the CCDs, which was determined by simulating events $<10~\mu$m from the front and back surfaces and adding them onto blank images.
This selection was validated over the full energy range in the surface laboratory by illuminating with a $^{14}$C $\beta$ source ($Q_\beta=156$~keV) the front and the back sides of a CCD from the same batch as those installed at SNOLAB.
The CCD was operated with the same parameters, readout settings and dark counts as in SNOLAB.
The leakage fraction of events following \sxy\ fiducial selection was 4$\pm$1\% and 3$\pm$1\% for backside and frontside illumination, respectively.

\begin{figure}[t]
	\centering
	\includegraphics[width=0.47\textwidth]{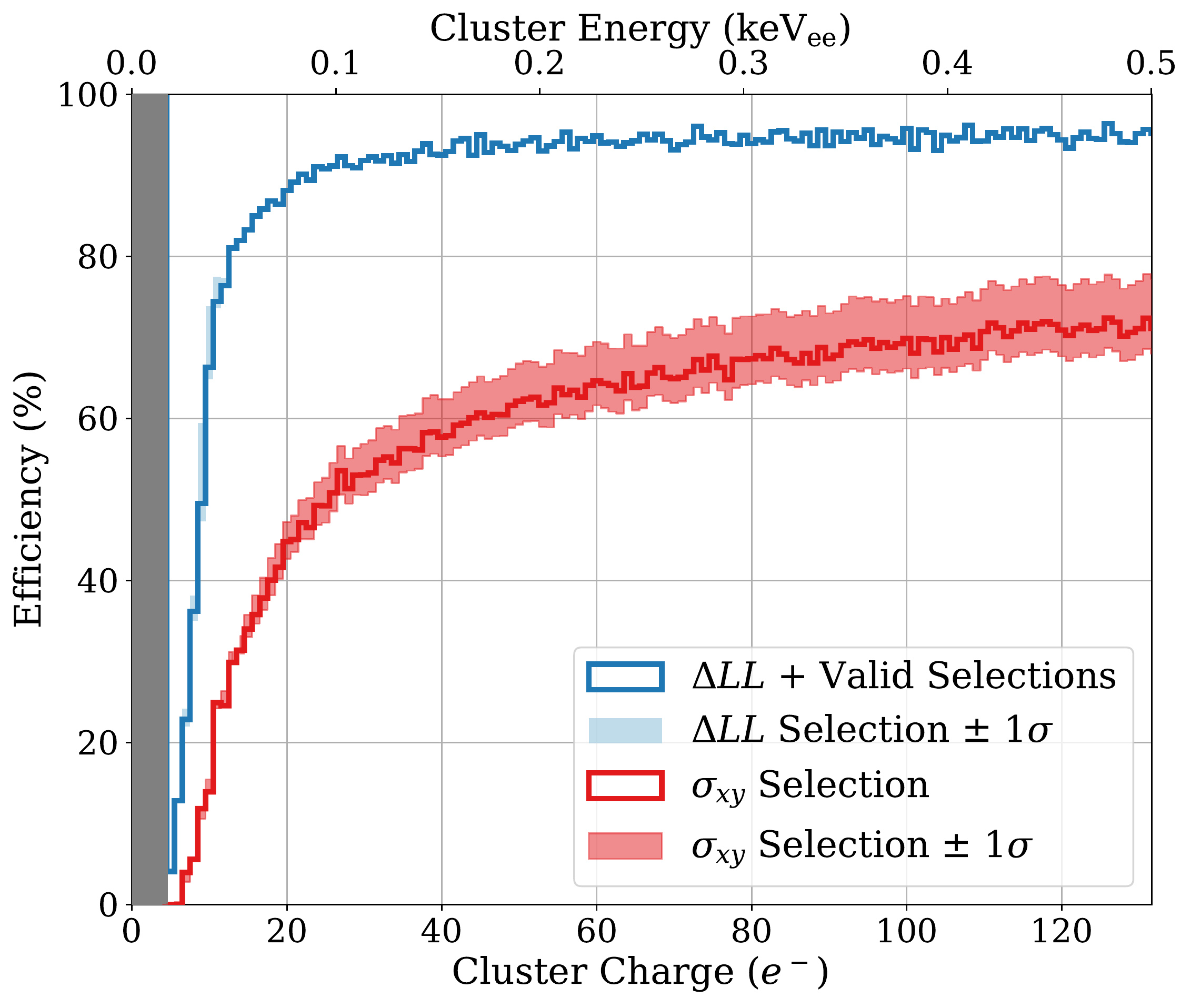}
	\caption{Selection efficiency for simulated bulk events after noise (blue line) and surface-event (red line) rejection with their respective uncertainty bands. The gray region below 5\,$e^-$ was not considered in this analysis.
}
	\label{fig:cut_efficiency}
\end{figure}

The selection efficiency for bulk events was obtained by simulating ionization events uniformly distributed in $z$ on blank images, applying the masks from data, and running the likelihood clustering.
Our diffusion model was previously validated for bulk events by irradiating a CCD with low-energy ($<24$\,keV) neutrons and $\gamma$ rays~\cite{Chavarria:2020ilz, PhysRevD.96.042002}.
Figure~\ref{fig:cut_efficiency} shows the efficiency of our selections as a function of energy.
After $\Delta LL$ and valid-cluster selections, we obtain $\geq 10\%$ efficiency for bulk events down to 6\,$e^-$ (23\,\eve ), which we set as the analysis threshold.
The efficiency plateaus at 95\% from the selection of valid clusters, with 4\% of clusters lost because the fit window contains at least one masked pixel.
The $\pm 1 \sigma$ uncertainty band is the statistical uncertainty in the number of selected events in the simulated blanks from the $\Delta LL$ selection.
For the $\sigma_{xy}$ selection, we estimated the 68\% confidence interval from simulations where the diffusion model parameters were varied within their uncertainties.

Although the copper box and CCD packages of the upgraded detector have a different geometry than in the previous installation, the apparatus components (\emph{i.e.}, CCDs, Kapton flex cables, glue, OFHC copper, brass screws and ancient lead), their composition and radioactivities remain nominally the same.
To confirm that the ionization spectrum features the same dominant background components as before, we first analyzed the spectrum above 0.5\,k\eve , where no excess above the background model was previously observed.
The overall (bulk) background rate in the 1--6\,k\eve\ energy range of $9.7\pm0.8$ ($4.4\pm0.6$)\,\dru\  is comparable to the previous installation~\cite{PhysRevLett.125.241803,*PhysRevD.105.062003}.
A spectral analysis up to 20\,k\eve\ showed the characteristic spectrum of cosmogenic tritium ($Q_\beta~=~18.6$~keV) in the bulk of the CCDs over an approximately constant background from Compton scattering of external $\gamma$ rays.
Surface events and x-ray lines from \pbten\ decays were also identified with comparable rates as before.

\begin{figure*}[htbp]
	\includegraphics[width=0.88\textwidth]{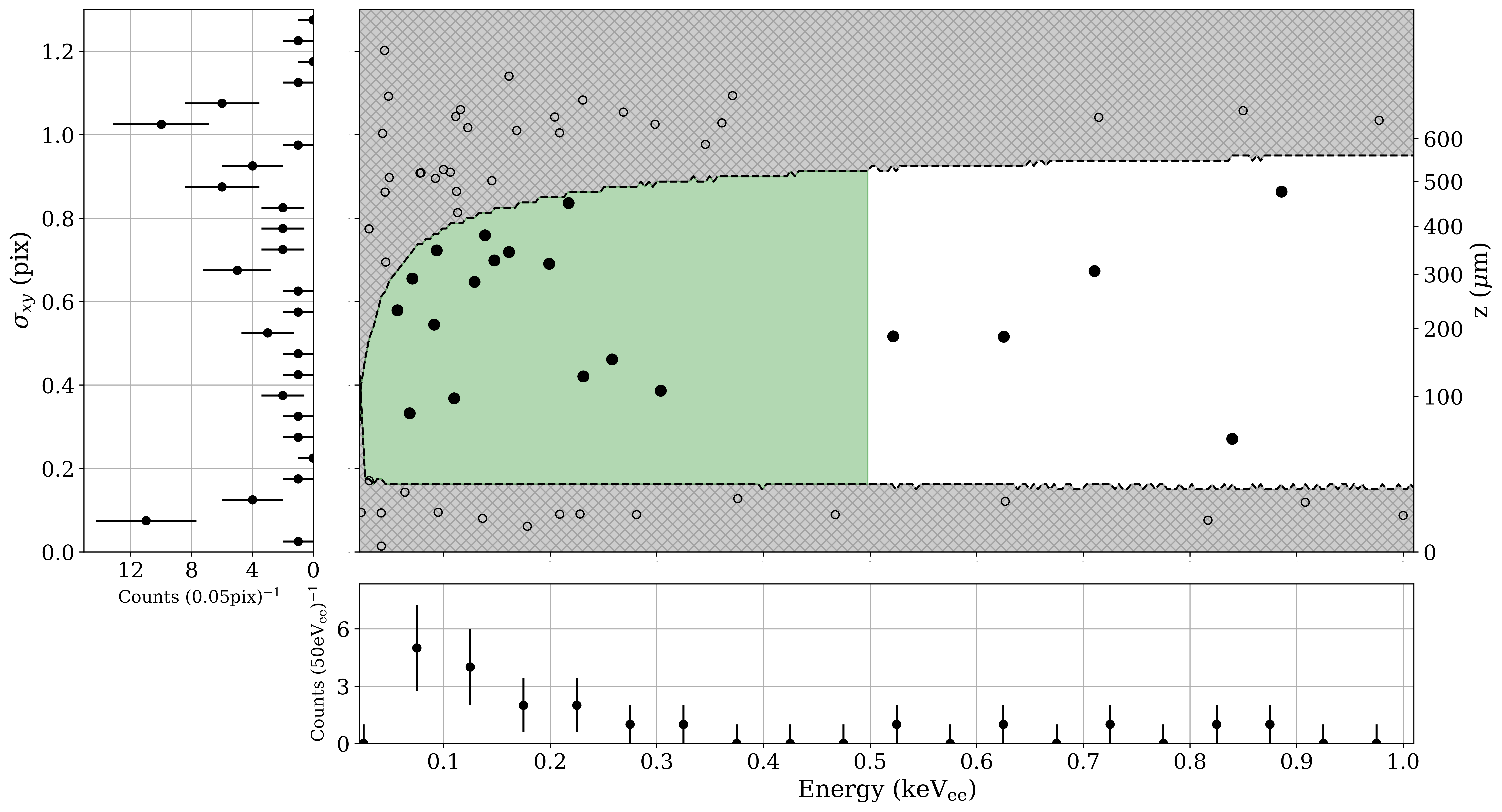}
	\caption{Distribution of clusters in energy vs. \sxy\ space, for $E\in[0.023,1]~$k\eve.
 Clusters with large (small) \sxy\ correspond to events occurring in the back (front) of the CCDs.
 Crosshatched gray regions are excluded by the 95\% surface event rejection (dashed black), which results in the final sample of fiducial events (filled circles). 
 The green-shaded region represents the energy ROI ($E<0.5~$k\eve) for the excess.
 The left (bottom) panel shows the \sxy\ (energy) projection of all (fiducial) events. 
 The depth ($z$) scale from the nominal diffusion model is reported on the right-hand side axis.
 }
	\label{fig:data_0_1000eV}
\end{figure*}

After establishing the final cluster selections and performing background studies, we unblinded the region of interest (ROI) below 0.5\,k\eve .
The clusters retained after the \dll\ and valid-cluster selections are shown in Fig.~\ref{fig:data_0_1000eV}, with energy and $\sigma_{xy}$ projections beside the corresponding axes.
The valid criteria remove 11 clusters that are contained in the fit window (with no masked pixels) and fall in the ROI.
The $\sigma_{xy}$ projection shows the distribution of all ionization events from threshold to 1.0~k\eve, while the energy projection shows the spectrum of fiducial events after \sxy\ selections, demarcated by the black dashed lines in the scatter plot.
Clusters in the ROI were individually inspected and were found to have the expected topology, located away from masked regions, and not spatially correlated with other ionization events.
The event $(x,y)$ and time distributions were confirmed to be statistically consistent with uniformity, with Kolmogorov-Smirnov p-values between 0.36 and 0.78.
The \sxy\ distribution was found to be consistent from a fit (p-value${=}0.73$) with three distinct populations of bulk, front- and back-surface events. A total of 15 events pass the fiducial \sxy\ selection in the ROI, in contrast with the 4.8$\pm$0.7  expected from the background rate measured in the 1--6\,k\eve\ range.
The increasing rate of events toward low energies at large \sxy\ is caused by partial charge collection of ionization events in the backside of the CCD, as described in Ref.~\cite{PhysRevLett.125.241803,*PhysRevD.105.062003}.

\begin{figure}[t]
	\centering
	\includegraphics[width=0.48\textwidth]{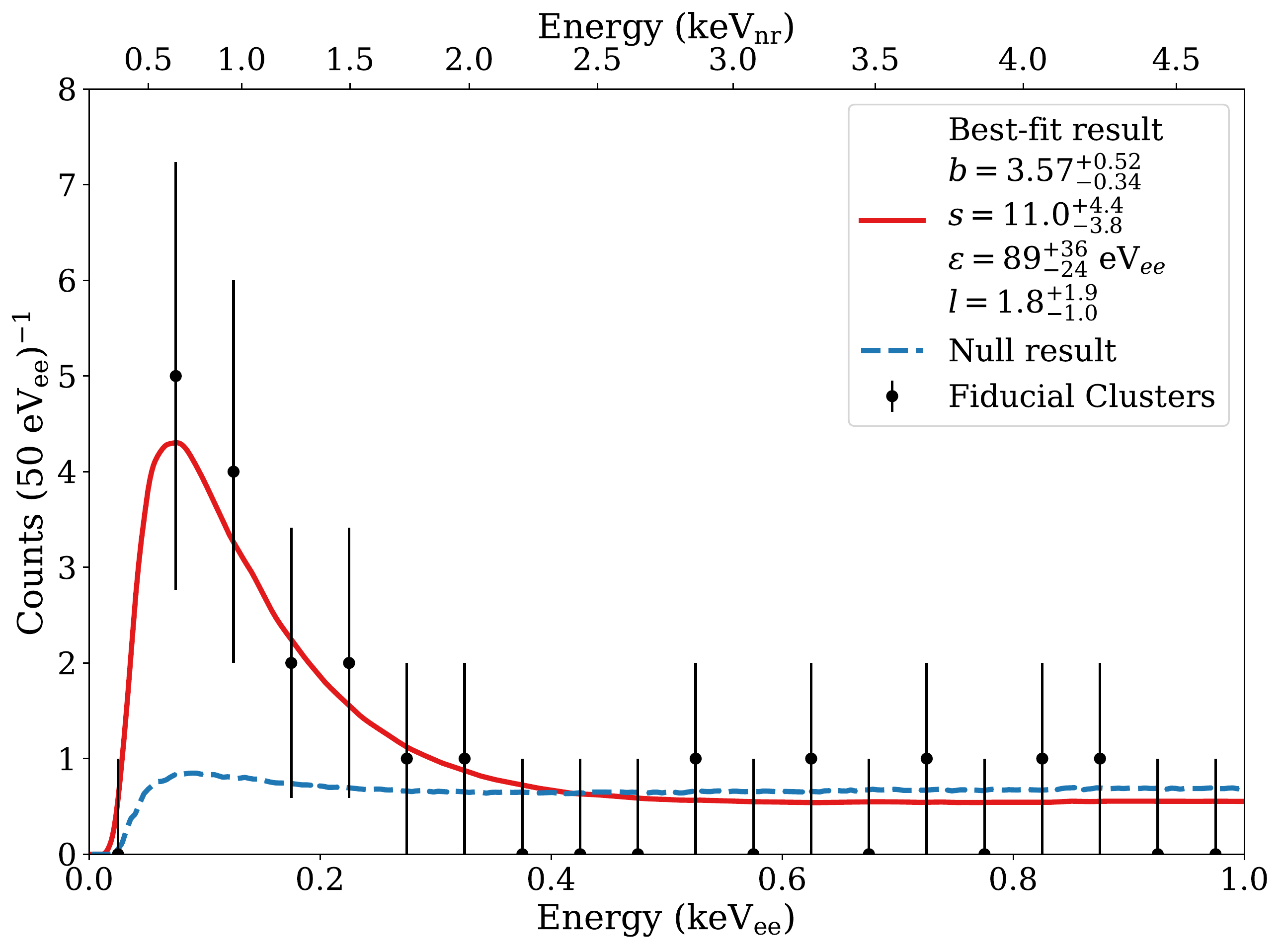}
	\caption{
 Result of the spectral fit to fiducial events with all parameters free (red) and the null hypothesis (blue dashed).
 The fit prefers an excess low-energy exponential component.
 Counts in the legend are reported for the ROI.
 The top axis shows the nuclear-recoil energy scale from Ref.~\cite{nrieyoussef}.}
	\label{fig:excess_energy_fit}
\end{figure}

\begin{figure}[t]
	\centering
	\includegraphics[width=0.5\textwidth]{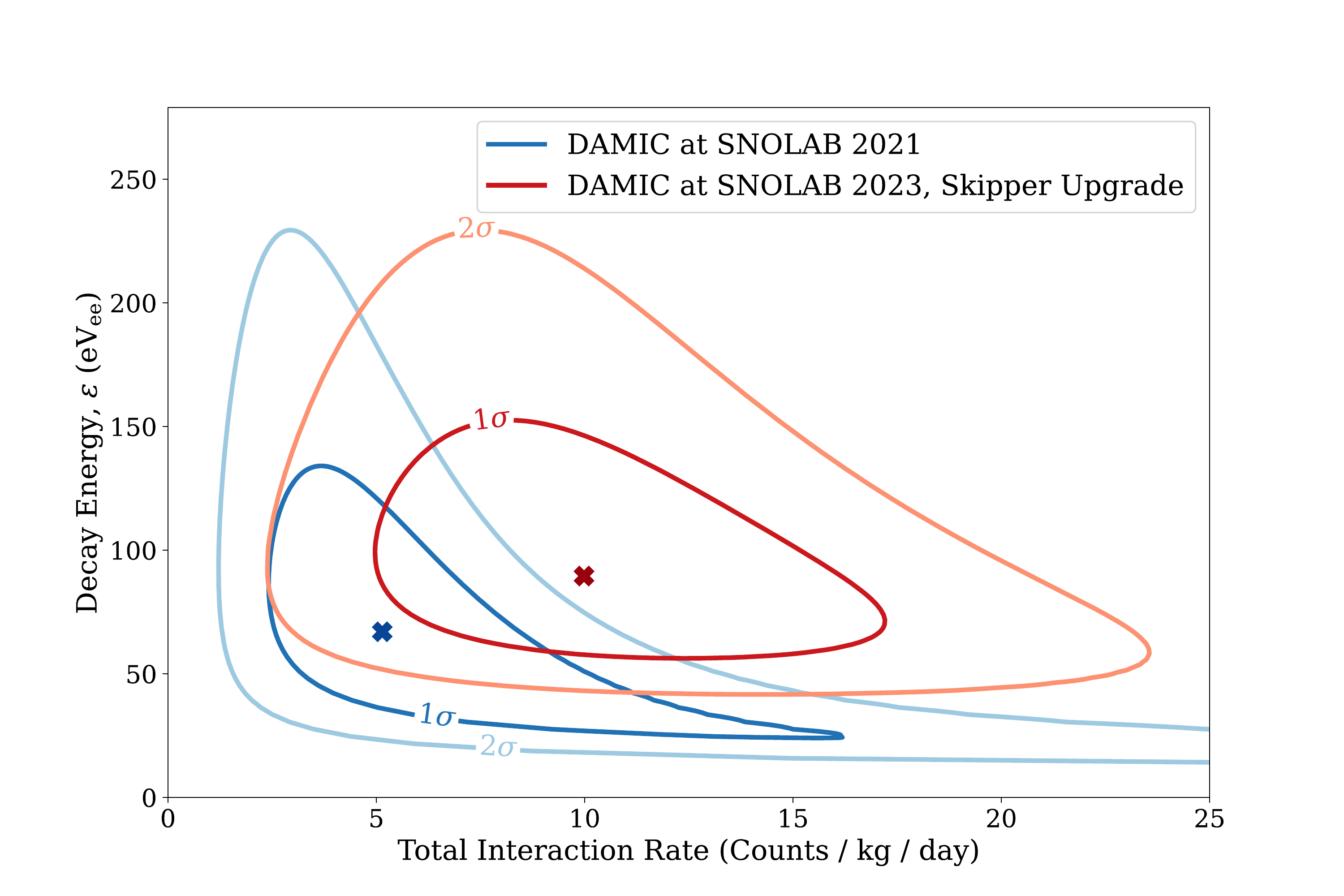}
	\caption{Comparison of the allowed parameter space for the bulk excess measured in this work and the previous 11\,\kgd\ analysis~\cite{PhysRevLett.125.241803,*PhysRevD.105.062003}.
 }
	\label{fig:excesses_contours}
\end{figure}

Our background model~\cite{PhysRevLett.125.241803,*PhysRevD.105.062003} predicts an approximately constant spectrum of bulk events at low energies.
This is expected since electronic recoils can only be induced in the bulk by $\beta$ decays or Compton scatterings, which both have approximately flat spectra in the ROI.\footnote{The Compton scattering spectrum in silicon is approximately flat between ${500}~$eV and ${200}~$eV, it drops by $50\%$ between ${200}~$eV and ${100}~$eV, and then plateaus~\cite{PhysRevD.106.092001}. This results in an up to 25\% decrease in the background rate below ${200}~$\eve , equivalent to only 5\% of the total spectrum. See Fig.~\ref{fig:excess_energy_fit}. }
Nuclear recoils from fast neutrons and thermal neutron captures are expected to have orders-of-magnitude lower rates~\cite{PhysRevLett.125.241803,*PhysRevD.105.062003}.
In addition, based on the number of surface (bulk) events rejected (retained) by the \sxy\ selection and the corresponding uncertainty from the diffusion model, we expect $1.8^{+1.9}_{-1.0}$ leakage events from the surface in the ROI.
We performed an extended unbinned likelihood fit to the energy spectrum of fiducial events with a flat background component and an exponentially decaying spectrum, which adequately parametrized the observed excess in the previous analysis.
The leakage is constrained to the expected value within uncertainty and assumed to contribute to the flat and exponential components as to have the same spectrum as the fiducial events in the ROI.
The free parameters in the fit are the integrated counts in the background ($b$), leakage ($l$) and excess ($s$) spectra, and the decay energy ($\varepsilon$) of the exponential.
The spectra were corrected for the bulk event acceptance (red line in Fig.~\ref{fig:cut_efficiency}) and the fit was performed between 0 and $6~\text{k}\eve$.
The results are shown in Fig.~\ref{fig:excess_energy_fit} for $E<1~$k\eve.
The best fit finds $s=11.0^{+4.4}_{-3.8}$ excess events with $\varepsilon=89^{+36}_{-24}~$\eve, and  $b=3.57^{+0.52}_{-0.34}$ background events in the ROI.
A likelihood-ratio test between the null hypothesis ($s=0$, with $b$ and $l$ free) and the best fit results in a p-value of $7.73 \times 10^{-4}$, corresponding to a significance of $3.4\sigma$.
We translate the number of fiducial excess events to a total rate of 10.0$^{+4.0}_{-3.4}$ per \kgd\ from the fiducial exposure.
The parameter space of the measured excess compared to the result from the previous 11 kg-day exposure~\cite{PhysRevLett.125.241803,*PhysRevD.105.062003} is shown in Fig.~\ref{fig:excesses_contours}.
The two results are statistically compatible and suggest a common origin of the excess population.

In summary, an excess population of bulk events above the expected background spectrum has been observed in two setups of DAMIC at SNOLAB, which employed CCDs with significantly different noise characteristics and readout modes.
Both CCDs with conventional~\cite{PhysRevD.96.042002} and skipper~\cite{PhysRevD.106.092001} readout have demonstrated the capability to correctly resolve spectral features at the energies where the excess is observed.
The topology of the events\textemdash  which can be resolved with high resolution thanks to the subelectron noise of skipper CCDs\textemdash is consistent with ionization events.
The modeling of surface backgrounds, which was the dominant systematic uncertainty in the previous analysis, has been addressed by analyzing a clean sample of bulk events selected using the lateral spread of the clusters (\sxy ).

The observed excess ionization events likely arise from an unidentified constant source of radiation in the DAMIC detector or from the environment, which is common to the two experiments.
As such, this excess is distinct from the excess of phonon signals reported in milli-Kelvin cryogenic calorimeters~\cite{Fuss:2022fxe, *Angloher:2022pas}, which are likely caused by stress released by the crystal~\cite{Anthony-Petersen:2022ujw}.
The only known interactions that could give rise to the observed excess spectrum are those from neutrons with silicon nuclei in the bulk of the CCDs.
The observed spectrum could be reproduced by the scattering of neutrons with energies up to {$\sim$17\,keV} and a flux of $\sim$0.2\,cm$^{-2}$d$^{-1}$ through the CCDs, but no such source of neutrons has been identified.
Turning to more exotic interpretations, the bulk excess spectrum is well described by nuclear recoils from interactions of weakly interacting massive particles (WIMPs).
For spin-independent WIMP-nucleus coherent elastic scattering with standard galactic halo parameters~\cite{recommended-conventions}, the excess corresponds to a WIMP with mass $\sim$2.5\,GeV/$c^2$ and a WIMP-nucleon scattering cross section $\sim$$3\times 10^{-40}$\,cm$^2$. This interpretation is nominally excluded by results from CDMSlite~\cite{SuperCDMS:2018gro} and DarkSide-50~\cite{DarkSide-50:2022qzh}.
Attempts to find a consistent interpretation between these experiments by systematically varying detector response  (\emph{e.g.}, nuclear-recoil ionization efficiencies), WIMP speed distribution, as well as alternate particle interaction models (\emph{e.g.}, Ref.~\cite{Feng:2011vu, *Feng:2013vaa}) are beyond the scope of this article.

The SENSEI experiment~\cite{SENSEI:2020dpa, *SENSEI:2023zdf}, currently taking data at SNOLAB, is designed to acquire a larger exposure of $\sim$40\,kg-day with skipper CCDs.
The DAMIC-M detector~\cite{DAMIC-M:2022aks}, a 0.7\,kg skipper-CCD array with an improved radioactive background rate of $\mathcal O(0.1)\,\dru$, is scheduled to start operations at the Modane Underground Laboratory in 2025.
If the bulk excess is detected in SENSEI and DAMIC-M, the significantly increased statistics will enable a high-resolution spectral measurement, studies of the time evolution of the excess, and investigations of its dependence on detector configuration and operating parameters to better understand its origin.

\begin{acknowledgments}
We are grateful to SNOLAB and its staff for support through underground space, logistical and technical services. SNOLAB operations are supported by the Canada Foundation for Innovation through the Major Science Initiatives Fund and the Province of Ontario Ministry of Colleges and Universities, with underground access provided by Vale at the Creighton mine site.
The DAMIC-M project has received funding from the European Research
Council (ERC) under the European Union's Horizon 2020 research and
innovation programme Grant Agreement No. 788137, and from the U.S. National Science Foundation (NSF) through
Grant No. NSF PHY-1812654.
We thank the College of Arts and Sciences at the University of Washington for contributing the first CCDs to the DAMIC-M project.
The CCD development at Lawrence Berkeley National Laboratory MicroSystems Lab was supported in part by the Director, Office of Science, of the U.S. Department of Energy under Contract No. DE-AC02-05CH11231.
The SENSEI Collaboration is grateful for the support of the Heising-Simons Foundation under Grant No. 79921.
We acknowledge financial support from the following agencies and organizations:
NSF through Grant No.\ NSF PHY-2110585 to the University of Washington and The University of Chicago; the Kavli Institute for Cosmological Physics at The University of Chicago through an endowment from the Kavli Foundation;
the U.S. Department of Energy Office of Science through the Dark Matter New Initiatives program;
Fermi National Accelerator Laboratory (Contract No. DE-AC02-07CH11359);
Institut Lagrange de Paris Laboratoire d'Excellence (under Reference No. ANR-10-LABX-63) supported by French state funds managed by the Agence Nationale de la Recherche within the Investissements d'Avenir program under Reference No. ANR-11-IDEX-0004-02;
Swiss National Science Foundation through Grant No. 200021\_153654 and via the Swiss Canton of Zurich;
IFCA through project PID2019-109829GB-I00 funded by MCIN/AEI;
Mexico's Consejo Nacional de Ciencia y Tecnolog\'{i}a (Grant No. CF-2023-I-1169) and  Direcci\'{o}n General de Asuntos del Personal Acad\'{e}mico--Universidad Nacional Aut\'{o}noma de M\'{e}xico (Programa de Apoyo a Proyectos de Investigaci\'{o}n e Innovaci\'{o}n Tecnol\'{o}gica Grant No. IN106322).
\end{acknowledgments}

\bibliography{myrefs.bib}

\end{document}